\documentclass[a4paper,11pt]{article}

\usepackage{amsmath,amsthm,array}
\usepackage{amssymb}
\usepackage{mathtools}
\usepackage{enumerate}
\usepackage[margin=3cm]{geometry}
 
\usepackage[utf8]{inputenc}
\usepackage{tgtermes}
\usepackage{mathptmx}
\usepackage[T1]{fontenc}
\usepackage{stackrel}
\usepackage{pict2e,picture}

\usepackage{hyperref}
\hypersetup{colorlinks=true}

\makeatletter
\newcommand{\pnrelbar}{%
  \linethickness{\dimen2}%
  \sbox\z@{$\m@th\prec$}%
  \dimen@=1.1\ht\z@
  \begin{picture}(\dimen@,.4ex)
  \roundcap
  \put(0,.2ex){\line(1,0){\dimen@}}
  \put(\dimexpr 0.5\dimen@-.2ex\relax,0){\line(1,1){.4ex}}
  \end{picture}%
}

\newcommand{\precneq}{\mathrel{\vcenter{\hbox{\text{\prec@neq}}}}}
\newcommand{\prec@neq}{%
  \dimen2=\f@size\dimexpr.04pt\relax
  \oalign{%
    \noalign{\kern\dimexpr.2ex-.5\dimen2\relax}
    $\m@th\prec$\cr
    \noalign{\kern-.5\dimen2}
    \hidewidth\pnrelbar\hidewidth\cr
  }%
}
\makeatother

\newtheorem{example}{Example}

\theoremstyle{definition}

\newcommand{\FF}{\mathbb{F}}

\newcommand{\QQ}{\mathbb{Q}}

\newcommand{\lcm}{\operatorname{lcm}}

\newcommand{\LT}{\operatorname{LT}}

\newcommand{\tp}{\operatorname{tp}}

\newcommand{\leftbimod}[3]{\vphantom{#1}^{#2}{\kern-#3pt #1}}

\numberwithin{equation}{subsection}
  
\title{A note on a paper by Hashemi and Kapur}

\author{Anna Nymann Heisel and Niels Lauritzen\thanks{Department of Mathematics, Aarhus Universitet, \href{mailto:annanheisel@gmail.com}{annanheisel@gmail.com}, 
\href{mailto:niels@math.au.dk}{niels@math.au.dk}.
}}
\begin{document}

\makeatletter
\let\@date\@empty
\makeatother
\maketitle 

\begin{abstract}
  Recently Hashemi and Kapur published an algorithm \cite{HK} for
  Gr\"obner basis conversion by truncating polynomials 
  according to a source and a target monomial order.
  Here we present a counterexample
  to this algorithm. 
\end{abstract}

\section*{Introduction}

Let $K$ be a field, $R = K[x_1, \dots, x_n]$ the polynomial ring in the 
variables $x_1, \dots, x_n$ over $K$ and $\langle S \rangle$ the ideal generated
by a subset $S\subset R$. Given two monomial orders 
$\prec_1$ and $\prec_2$ on $R$, Hashemi and Kapur defined in \cite[Definition 3.1]{HK} the 
truncated part $\tp_{1, 2}(f)$ of a non-zero polynomial $f\in R$ as 
the sum of the terms in $f$ greater than or equal to $\LT_{\prec_1}(f)$
with respect to $\prec_2$. 

\begin{example}
If $R = K[x, y, z]$ and $\prec_1$ is the degree reverse lexicographic order given 
by $z \prec_1 y \prec_1 x$ and $\prec_2$ is the lexicographic order given by 
$z \prec_2 y \prec_2 x$, then 
$$
\tp_{1, 2}(f) = y^2 + x z + x
$$
for $f = y^2 + x z + x + y + z$.
\end{example}
 
If $G$ is a finite subset of polynomials containing 
a Gr\"obner basis with respect to $\prec_1$ for $\langle G \rangle$, the idea in \cite{HK} is 
to compute a Gr\"obner basis for $\langle G \rangle$ over $\prec_2$, by successive 
G\"obner basis computations over $\prec_2$ for $\tp_{1, 2}(G)$ 
with lifting steps as in the classical
Gr\"obner walk algorithm. Basically one discards $S$-polynomials of truncated polynomials 
in $\tp_{1, 2}(G)$, which reduce
to zero and lifts the non-zero reductions by keeping track of 
coefficients in the division algorithm (Gr\"obner coefficients). We refer to 
\cite[\S 4]{HK} for details.  

Difficulties understanding the proof of \cite[Theorem 4.5]{HK} prompted 
us to look for counterexamples.
Computer experiments by the first author over the rational numbers 
\cite[\S 3.4]{NH}\footnote{We thank Anders Nedergaard Jensen for double-checking this first counterexample.} finally revealed  
that \cite[Algorithm 2]{HK} is incorrect as stated. In this note 
we give an example over the field $K=\FF_2$ with two elements, simple enough for hand calculation, 
showing that the algorithm fails to produce a Gr\"obner basis 
with respect to $\prec_2$. This example was found using the Sage (pythonic) interface to Macaulay2.

\section{A counterexample over the field with two elements}

Let $R = \FF_2[x, y, z]$ .
We let $\prec_1$ be the degree reverse lexicographic order 
on $R$ with 
$z \prec_1 y \prec_1 x$ and $\prec_2$ 
the lexicographic order with        
$z \prec_2 y \prec_2 x$.

Let
\begin{align*}
    g_1 &= y^2 + x z + x\\
    g_2 &= z^2 + 1.
\end{align*}
Since $\LT_{\prec_1}(g_1) = y^2$ and $\LT_{\prec_1}(g_2) = z^2$ are relatively prime, $G_1 = \{g_1, g_2\}$
 is the reduced Gr\"obner basis
with respect to $\prec_1$ for the ideal $I$ generated by $G_1$.

The reduced Gr\"obner basis $G_2$ with respect to $\prec_2$ for $I$ is found by adding the polynomials
\begin{align}\label{countertwo}
\begin{split}
&y^2 z + y^2 \\
&y^4.
\end{split}
\end{align}

The steps of \cite[Algorithm 2]{HK} for $G_1$ can be carried out by hand without excessive computations.
We do this below.

\subsection{Computations}

\subsubsection{\cite[Algorithm 2, line 3]{HK}}
\noindent
$\tp_{1, 2}(G_1)$ is $F = \{h_1, h_2\}$ with
\begin{align*}
h_1 &= x z + x + y^2\\
h_2 &= z^2
\end{align*}

\subsubsection{\cite[Algorithm 2, line 4 and Algorithm 1 with line 13]{HK}}
Here
$$
S(h_1, h_2) = z h_1 + x h_2 = x z + y^2 z = h_1 + x + y^2 z + y^2
$$
with Gr\"obner coefficient $[z+1, x]$. Adding $h_3:=x + y^2 z + y^2$ to $F$ gives a
Gr\"obner basis $G$ for $\tp_{1, 2}(G_1)$ with respect to $\prec_2$, since 
$S(h_1, h_3)$ and $S(h_2, h_3)$ have zero remainders modulo $\{h_1, h_2, h_3\}$.

\subsubsection{\cite[Algorithm 2, lines 6, 7]{HK}}

The Gr\"obner coefficient $[z+1, x]$ gives the lifting
$$
g_3 = (z + 1)g_1 + x g_2 = y^2 z + y^2,
$$
which is added to $G_2$. 

\subsubsection{\cite[Algorithm 2, line 8]{HK}}

The interreduction of $G_2$ is ($G_2$ itself)
\begin{equation}\label{truncstop}
G_2 = \{g_1, g_2, g_3\}.
\end{equation}

\subsubsection{\cite[Algorithm 2, lines 9 and 10]{HK}}\label{step}

Here $\tp_{1, 2}(G_2)$ is $H = \{h_1, h_2, h_3\}$ with
\begin{align*}
    h_1 &=x z + x + y^2 \\
    h_2 &=z^2, \\
    h_3 &= y^2 z.
\end{align*}


The least common multiple of the leading terms of $h_1$ and $h_2$ is $x z^2$ and 
the least common multiple of the leading terms of $h_1$ and $h_3$ is $x y^2 z$.
According to \cite[Algorithm 1, line 6]{HK}, the $S$-polynomial $S(h_1, h_2)$ (with the minimal $\lcm$)
is processed first. So
$$
S(h_1, h_2) = z h_1 + x h_2 = x z  + y^2 z = h_1 + h_3 + x + y^2.
$$
and $h_4 := x + y^2$ with Gr\"obner coefficient  $[z+1, x, 1]$ is added to $H$. 
Therefore $S(h_1, h_3) = y^2 h_1 + x h_3 = x y^2 + y^4$ reduces to zero and 
is not added.
One verifies that 
$$
S(h_1, h_4) = h_1 + z h_4 = y^2 z + y^2 + z = h_3 + y^2 + z
$$
so that $h_5 := y^2 + z$ gets added with Gr\"obner coefficient $[1, 0, 1, z]$. Finally all
$S$-polynomials of $\{h_1, h_2, h_3, h_4, h_5\}$ reduce to zero.

\subsubsection{\cite[Algorithm 2, lines 6 and 7]{HK}}

The Gr\"obner coefficent $[z+1, x, 1]$ gives the lifting
$$
g_4 = (z+1) g_1 + x g_2 + g_3 = 0.
$$
The Gr\"obner coefficient $[1, 0, 1, z]$ gives the lifting
$$
g_5 = g_1 + g_3 + z g_4  = x z + x + y^2 z.
$$

The interreduction of $\{g_1, g_2, g_3, g_4, g_5\}$ is $\{g_1, g_2, g_3\}$, which 
we observed in \eqref{truncstop}
and the 
algorithm terminates by \cite[Algorithm 2, line 11]{HK} with $\{g_1, g_2, g_3\}$ as output missing the 
polynomial $y^4$ in the actual Gr\"obner basis, cf.~\eqref{countertwo}.

\section{Remarks} 

Had we chosen to process $S(h_1, h_3)$ before $S(h_1, h_2)$ in \S \ref{step}, we would have
found $y^4$ in the Gr\"obner basis ending up with the correct result.
A more complicated counterexample \cite[A.1, \texttt{example 4}]{NH} is given by
$\{x_1^3 x_2^5 x_3 + x_1^3 x_3, 
x_1^7 x_2 + x_1^3 + x_2 x_3 x_4, 
x_1^7 x_3 - x_1^3 x_2^4 x_3 - x_2^5 x_3^2 x_4, 
x_2^6 x_3 + x_2 x_3, 
x_4^3 + x_1\} \subseteq K[x_1, x_2, x_3, x_4]$ for 
$K = \FF_2, \QQ$ and $x_4\prec \cdots \prec x_1$ with $\prec = \prec_1, \prec_2$
as above. Here no permutation of the $S$-polynomials 
gives the correct result in the extended Buchberger algorithm 
in \cite[Algorithm 2]{HK} with the standard implementation of
the division algorithm.

\newcommand{\germ}{\mathfrak}

\providecommand{\bysame}{\leavevmode\hbox to3em{\hrulefill}\thinspace}
\providecommand{\MR}{\relax\ifhmode\unskip\space\fi MR }
\providecommand{\MRhref}[2]{%
  \href{http://www.ams.org/mathscinet-getitem?mr=#1}{#2}
}
\providecommand{\href}[2]{#2}



\end{document}